# An information upper bound for probability sensitivity

Jiannan Yang*

*Department of Engineering, University of Cambridge, Trumpington Street, Cambridge CB2 1PZ, UK*

Uncertain input of a mathematical model induces uncertainties in the output and probabilistic sensitivity analysis identifies the influential inputs to guide decision-making. Of practical concern is the probability that the output would, or would not, exceed a threshold, and the probability sensitivity depends on this threshold which is often uncertain. The Fisher information and the Kullback-Leibler divergence have been recently proposed in the literature as threshold-independent sensitivity metrics. We present mathematical proof that the information-theoretical metrics provide an upper bound for the probability sensitivity. The proof is elementary, relying only on a special version of the Cauchy-Schwarz inequality called Titu's lemma. Despite various inequalities exist for probabilities, little is known of probability sensitivity bounds and the one proposed here is new to the present author's knowledge. The probability sensitivity bound is extended, analytically and with numerical examples, to the Fisher information of both the input and output. It thus provides a solid mathematical basis for decision-making based on probabilistic sensitivity metrics.

## 1 Introduction

The use of mathematical models to simulate real world phenomena is firmly established in many areas of science and technology. Such models are typically implemented in computer programs and the modelling process often involves synthetizing data from multiple sources and of different level of relevance. The influence of these uncertain data on the model output most often is not transparent, and sensitivity analysis thus becomes an indispensable part of the modelling process.

A broad range of approaches can be found in the literature, but in practice the input uncertainties are commonly quantified by a joint probably distribution. A suitable measure, such as the expected value or variance, can then be used to summarise the induced output uncertainties. More commonly in practice, the concern is often the probability that the output would (not) exceed a certain threshold. For example, the probability of cost-effectiveness in health economics [1,2], the probability of acceptability in design

*E-mail address*: jy419@cam.ac.uk (J. Yang).

[3] and the probability of failure in reliability engineering [4,5] and material design [6]. The probabilistic sensitivity analysis then examines the relationship between the uncertain input and the induced uncertainty of the output [7]. In particular, in the context of decision making, an important task is to identify the uncertain input that influence the acceptance/failure probability the most.

In this paper, we present a novel upper bound for the sensitivity of the acceptance/failure probability using the Fisher information and the Kullback-Leibler (K-L) divergence. While there are various probability inequalities and bounds, such as the Chebyshev's inequality that provides an upper bound on the tail probability of a random variable based on its mean and variance [8], little is known of probability sensitivity bounds and the one proposed here is new to the present author's knowledge.

The main motivation of deriving this bound is that the probability sensitivity depends on a (type of) threshold, and that is often evolving during a decision process. One implication for the evolving threshold is that, the acceptance function or failure mode of interest can change as the understanding of the problem evolves and this is especially true for processes where decisions are made under large uncertainties, e.g. early concept design phase. To capture this evolving uncertainties, in a recent study, the Fisher information have been proposed as a new sensitivity metric that is independent of this evolving threshold [9]. It was demonstrated in [9], via numerical case studies, that the sensitivity information extracted from the Fisher information is consistent with the probability sensitivities which depend on the decision thresholds. This close relationship between the two metrics is expected to provide a consistent identification of high value information throughout the design process. Therefore, the sensitivity inequality presented in this paper, where the probability sensitivity is found to be bounded by the thresholds-independent Fisher information, would provide a solid mathematical basis to the aforementioned design tools, and more generally to decision making under uncertainties. In addition to the application to decision makings, it is envisaged that the probability sensitivity bound would make contribution to stochastic optimizations [10]. In many applications in system reliability, operational research and financial optimization, the objective is often to minimize the probability of failure under constraints. In these stochastic optimization problems, either or both of the objective function and the constraint are defined probabilistically and the probability sensitivity/gradient is an essential element of gradient based optimizations.

In what follows, the mathematical proof of the probability sensitivity bound is given in details in *Section 2.* Moreover, we show that the first order perturbation of the probability is never bigger than the change of entropy defined using the Kullback-Leibler divergence. In *Section 3*, an intuitive view of the bounds is first presented. Next, the sensitivity bound is extended to the Fisher information of both the input and output using the information processing inequality. Lastly, some numerical considerations for an efficient implementation of the sensitivity analysis are given in the same section. In *Section 4*, several examples are given to demonstrate the validity of the sensitivity bound. Concluding remarks are given in *Section 5*.

## 2 Mathematical proof

In this section, we demonstrate that there is an upper bound on the sensitivity of a probability measure:

$$\left\|\frac{\partial P_{\mathrm{f}}(\mathbf{b},z)}{\partial \mathbf{b}}\right\|^2 \le \mathrm{tr}\{\mathbf{F}(\mathbf{b})\} \tag{1}$$

where $\mathbf{b}$ is the parameter of the uncertain inputs and $P_{\mathrm{f}}$ is the failure (or acceptance) probability that depends on the threshold $z$. The matrix $\mathbf{F}$ is the symmetric semi-positive definite Fisher Information Matrix (FIM) [11]. In addition, we also present a bound that links the first order perturbation of the probability to entropy:

$$\left|\Delta P_{\mathrm{f}}(\mathbf{b},z)\right|^2 \le \Delta\mathbf{b}^{\mathrm{T}}\mathbf{F}\Delta\mathbf{b} = 2\Delta H \tag{2}$$

where $\Delta H$ is the relative entropy defined in *Eq (3)* and *(4)*.

### *2.1 Preliminaries*

Consider a general mapping from the input $\mathbf{x}$ to the output $\mathbf{y}$, where the function $\mathbf{y} = \mathbf{h}(\mathbf{x})$ is deterministic. It is assumed that $\mathbf{x}$ is random and its uncertainties can be described by some probability distribution $p(\mathbf{x})$ that is dependent on parameters $\mathbf{b}$ (e.g., $b$ is the distribution parameter). The interest is often to characterise the uncertainties of the output $\mathbf{y}$ that is induced by the input $\mathbf{x}$. When the joint probability distribution of the output is known, the entropy of the uncertainty can be estimated as:

$$H = -\int p(\mathbf{y}\,|\,\mathbf{b}) \ln p(\mathbf{y}\,|\,\mathbf{b}) \mathrm{d}\mathbf{y} \tag{3}$$

The perturbation of the entropy is defined as a relative entropy quantified using the K-L divergence:

$$\Delta H \equiv KL\left[p(\mathbf{y}\,|\,\mathbf{b}) \,\|\, p(\mathbf{y}\,|\,\mathbf{b}+\Delta\mathbf{b})\right] = \int p(\mathbf{y}\,|\,\mathbf{b}) \ln\left[\frac{p(\mathbf{y}\,|\,\mathbf{b})}{p(\mathbf{y}\,|\,\mathbf{b}+\Delta\mathbf{b})}\right]\mathrm{d}\mathbf{y} \approx \frac{1}{2}\Delta\mathbf{b}^{\mathrm{T}}\mathbf{F}\Delta\mathbf{b} \tag{4}$$

where the third and higher order terms from the Taylor expansion of the perturbed probability are ignored at the last step (see *Appendix A* for more details). The *jk*th entry of the Fisher Information Matrix (FIM) can be expressed as:

$$F_{jk} = \int \frac{\partial p(\mathbf{y})}{\partial b_j}\frac{\partial p(\mathbf{y})}{\partial b_k}\frac{1}{p(\mathbf{y})}\mathrm{d}\mathbf{y} = \mathbb{E}_{\mathbf{Y}}\left[\frac{\partial \ln p(\mathbf{y})}{\partial b_j}\frac{\partial \ln p(\mathbf{y})}{\partial b_k}\right] \tag{5}$$

As mentioned in the introduction, one summary measure commonly used in practice is the probability that the output would exceed a certain threshold:

$$P_{\mathrm{f}}(\mathbf{b},z) = \int \mathrm{H}\left(g(\mathbf{y}) - z\right) p(\mathbf{y}\,|\,\mathbf{b}) \mathrm{d}\mathbf{y} \tag{6}$$

where $\mathrm{H}(\cdot)$ is the Heaviside step function and $z$ represents the threshold. The function $g(\cdot)$ is the performance function or mode of failure. When the function $g(\cdot)$ is an identity function, the resulted failure probability is equivalent to the (Complementary) Cumulative Distribution Function (CDF) of the random output $\mathbf{y}$. Nevertheless, the function $g(\cdot)$ is a transformation of the output and can be any nonlinear function in general. The first order perturbation of the failure probability is then given as:

$$P_{\mathrm{f}}\left(\mathbf{b}+\Delta\mathbf{b},z\right)=P_{\mathrm{f}}\left(\mathbf{b},z\right)+\Delta P_{\mathrm{f}}\ \approx P_{\mathrm{f}}\ \left(\mathbf{b},z\right)+\frac{\partial P_{\mathrm{f}}\left(\mathbf{b},z\right)}{\partial\mathbf{b}}\Delta\mathbf{b} \tag{7}$$

where

$$\frac{\partial P_{\mathrm{f}}\left(\mathbf{b},z\right)}{\partial\mathbf{b}}=\begin{bmatrix}\frac{\partial P_{\mathrm{f}}}{\partial b_1} & \frac{\partial P_{\mathrm{f}}}{\partial b_2} & \cdots & \frac{\partial P_{\mathrm{f}}}{\partial b_n}\end{bmatrix} \tag{8}$$

is a row vector of the partial derivatives.

Assuming the differential and integral operators are commutative, i.e. the order of the two operations can be exchanged under regularity conditions of continuous and bounded functions [10], we have the sensitivity/gradient as:

$$\frac{\partial P_{\mathrm{f}}\left(\mathbf{b},z\right)}{\partial\mathbf{b}}=\int\mathrm{H}\left(g\left(\mathbf{y}\right)-z\right)\frac{\partial p\left(\mathbf{y}\,|\,\mathbf{b}\right)}{\partial\mathbf{b}}\mathrm{d}\mathbf{y} \tag{9}$$

Note that the failure probability and its sensitivity can also be estimated directly from the joint probability density function of the input, as shown in *Eq (25)* and *(34)* in *Section 3*.

## 2.2 Sensitivity bound

In this section, we give proof of the sensitivity bound in *Eq (1)*. Starting from *Eq (9)* and consider the squared amplitude of the $k^{\text{th}}$ term of the probability gradient:

$$\left(\frac{\partial P_{\mathrm{f}}}{\partial b_k}\right)^2=\left[\int\mathrm{H}\left(g\left(\mathbf{y}\right)-z\right)\frac{\partial p\left(\mathbf{y}\,|\,\mathbf{b}\right)}{\partial b_k}\mathrm{d}\mathbf{y}\right]^2\le\left[\int\left|\frac{\partial p\left(\mathbf{y}\,|\,\mathbf{b}\right)}{\partial b_k}\right|\mathrm{d}\mathbf{y}\right]^2 \tag{10}$$

where the triangle inequality is used at the last step and note that the amplitude of the Heaviside function is smaller than one by definition, i.e. $|\mathrm{H}(\cdot)|\le 1$. Divide the last term in *Eq (10)* by one:

$$\left[\int\left|\frac{\partial p\left(\mathbf{y}\,|\,\mathbf{b}\right)}{\partial b_k}\right|\mathrm{d}\mathbf{y}\right]^2=\frac{\left[\int\left|\frac{\partial p}{\partial b_k}\right|\mathrm{d}\mathbf{y}\right]^2}{\int p\mathrm{d}\mathbf{y}}\le\int\left|\frac{\partial p}{\partial b_k}\right|^2\frac{1}{p}\mathrm{d}\mathbf{y} \tag{11}$$

where the first equality assumed the joint PDF is normalised and has unit area, i.e. $\int p\mathrm{d}\mathbf{y}=1$. For the inequality at the last step in *Eq (11)*, a special version of the Cauchy-Schwarz inequality is applied, also known as Titu's lemma or Sedrakayan's inequality [12]. The Titu's lemma states that for positive real numbers $u_i$ and $v_i$, the following inequality holds:

$$\frac{\left(\sum_{i=1}^{n} u_i\right)^2}{\sum_{i=1}^{n} v_i} \leq \sum_{i=1}^{n} \frac{u_i^2}{v_i} \tag{12}$$

and this leads naturally to the integral inequality used above. From *Eq (11)*, the L2 norm of the probability sensitivity vector can be found as:

$$\left\| \frac{\partial P_{\mathrm{f}}}{\partial \mathbf{b}} \right\|^2 = \sum_k \left( \frac{\partial P_{\mathrm{f}}}{\partial b_k} \right)^2 \leq \sum_k \int \left| \frac{\partial p}{\partial b_k} \right|^2 \frac{1}{p} \mathrm{d}\mathbf{y} = \mathrm{tr}\left(\mathbf{F}\right) \tag{13}$$

where the proof for *Eq (1)* is completed.

### 2.3 Perturbation bound

The sensitivity bound presented in the previous section makes no approximations except the regularity conditions so that the order of the integral and differential operations can be exchanged [10]. In this section, we show that the first order perturbation of the probability is bounded by the relative entropy. Note that this is only valid for infinitesimal change of the parameter **b**.

The squared amplitude of the first order perturbation of the failure probability, from *Eqs (7)* and *(9)*, can be expressed as:

$$\left|\Delta P_{\mathrm{f}}\right|^2 = \left| \frac{\partial P_{\mathrm{f}}}{\partial \mathbf{b}} \Delta \mathbf{b} \right|^2 = \left| \int \mathrm{H}\left(g\left(\mathbf{y}\right) - z\right) \frac{\partial p}{\partial \mathbf{b}} \Delta \mathbf{b} \mathrm{d}\mathbf{y} \right|^2 \tag{14}$$

where $\partial p / \partial \mathbf{b}$ is defined as a row vector as in *Eq (8)*. Applying Titu's lemma and noting that $|\mathrm{H}(\cdot)| \leq 1$ and $\int p \mathrm{d}\mathbf{y} = 1$ as in *Eqs (10)* and *(11)*, we have the following inequality:

$$\left| \int \mathrm{H}\left(g\left(\mathbf{y}\right) - z\right) \frac{\partial p}{\partial \mathbf{b}} \Delta \mathbf{b} \mathrm{d}\mathbf{y} \right|^2 \leq \left( \int \left| \frac{\partial p}{\partial \mathbf{b}} \Delta \mathbf{b} \right| \mathrm{d}\mathbf{y} \right)^2 \leq \int \left| \frac{\partial p}{\partial \mathbf{b}} \Delta \mathbf{b} \right|^2 \frac{1}{p} \mathrm{d}\mathbf{y} \tag{15}$$

Expand the squared term and note the definition of the FIM, we then have following relationship:

$$\int \left| \frac{\partial p}{\partial \mathbf{b}} \Delta \mathbf{b} \right|^2 \frac{1}{p} \mathrm{d}\mathbf{y} = \Delta \mathbf{b}^{\mathrm{T}} \left( \int \frac{\partial p}{\partial \mathbf{b}} \frac{\partial p}{\partial \mathbf{b}}^{\mathrm{T}} \frac{1}{p} \mathrm{d}\mathbf{y} \right) \Delta \mathbf{b} = \Delta \mathbf{b}^{\mathrm{T}} \mathbf{F} \Delta \mathbf{b} \tag{16}$$

Noting *Eq (4)*, the perturbation of the failure probability is found to be bounded by the relative entropy:

$$\left|\Delta P_{\mathrm{f}}\right|^2 \leq \Delta \mathbf{b}^{\mathrm{T}} \mathbf{F} \Delta \mathbf{b} = 2 \Delta H$$

where the proof for *Eq (2)* is completed. Note that this bound is only effective when $\Delta H \leq 1/2$ because $|\Delta P_{\mathrm{f}}|^2 \leq 1$.

Although we make no assumption of the form of the function $g(\cdot)$, when it is an identity function in *Eq (6)*, $P_{\mathrm{f}}$ is simply the Cumulative Distribution Function. In this special case, the perturbation bound is a consequence of the Pinsker's inequality [13]. One form of the Pinsker's inequality relates the L1 norm between two discrete probability distribution, $P$ and $Q$, to the K-L divergence [14]:

$$\|P-Q\|_1^2 \le 2KL\left[P \,\|\, Q\right] \tag{17}$$

If we assume the distribution $P$ is the perturbation of the distribution $Q$, i.e. $Q = P_{\mathrm{f}}(\mathbf{b}) = \sum_{i\in A} p_i(\mathbf{b})$ where $A$ is the failure region, and $P = P_{\mathrm{f}}(\mathbf{b}+\Delta\mathbf{b}) = \sum_{i\in A} p_i(\mathbf{b}+\Delta\mathbf{b})$ where $p_i$ is the probability mass:

$$\left|\Delta P_{\mathrm{f}}\right|^2 = \left|\sum\nolimits_{i\in A} p_i(\mathbf{b}+\Delta\mathbf{b}) - p_i(\mathbf{b})\right|^2 \le \left(\sum\nolimits_{i\in A}\left|p_i(\mathbf{b}+\Delta\mathbf{b}) - p_i(\mathbf{b})\right|\right)^2 = \|P-Q\|_1^2 \tag{18}$$

where the bound in *Eq (2)* follows immediately using Pinsker's inequality in *Eq (17)*.

# 3 Discussion

## *3.1 A geometric view of the bound*

In this section, we give an intuitive geometric view of the sensitivity bound using the perturbation inequality in *Eq (2)* as an example. For simplicity, we assume the function $g(\cdot)$ is an identity function in *Eq (6)* and the resulted failure probability is simply the Cumulative Distribution Function (CDF) of the random output $\mathbf{y}$. Consider a probability mass function (PMF) in $\mathbb{R}^d$ with its simplex representation [11,15]:

$$\sum_i p_i = 1,\ p_i = p\left(y_i\right) \ge 0; \text{ for } i = 1,\ldots,d \tag{19}$$

with a change of variable, $q_i = \sqrt{p_i}$, we have

$$\sum_i q_i^2 = \sum_i p_i = 1; \text{ for } i = 1,\ldots,d \tag{20}$$

which describes the L2 norm of the vector $\mathbf{q} = [q_i, q_i, \ldots, q_d]$ in $\mathbb{R}^d$. Therefore, all PMF in $\mathbb{R}^2$ can be represented as a surface of a circle with unit radius in the positive quadrant, as seen in *Figure 1*.

For failure probabilities:

$$P_{\mathrm{f}} = \sum_{i\in A} p_i = \sum_{i\in A} q_i^2 \tag{21}$$

where $A$ is the failure region. Taking the vector view described above, the failure probability can be seen as a subspace projection of the vectors $\mathbf{q}$. Therefore, we have the following inequality:

$$\left|D_2\right|^2 = \left|\Delta P_{\mathrm{f}}\right|^2 = \left|P_{\mathrm{f}}\left(\mathbf{b}+\Delta\mathbf{b}\right) - P_{\mathrm{f}}\left(\mathbf{b}\right)\right|^2 \le \left|D_1\right|^2 \tag{22}$$

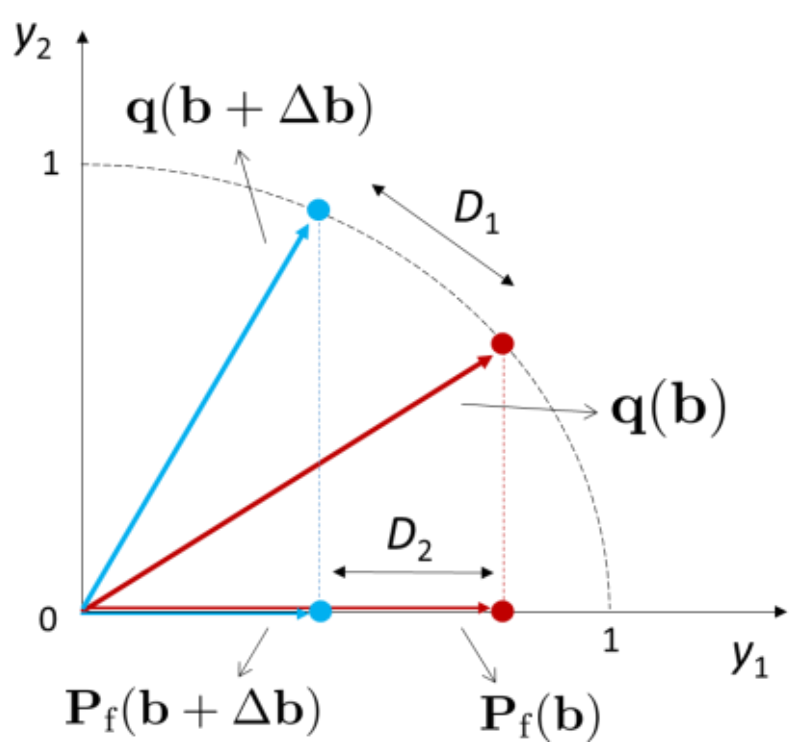

*Figure 1 A geometric view of the probablity perturbation bound*

The distance $D_1$ due to a change of the parameter $\mathbf{b}$ can be computed as:

$$|D_1|^2 = \sum_i \left| q_i(\mathbf{b}+\Delta\mathbf{b}) - q_i(\mathbf{b}) \right|^2 \tag{23}$$

Considering first order perturbation as in *Eq (7)*, i.e. $q_i(\mathbf{b}+\Delta\mathbf{b}) = q_i(\mathbf{b}) + \partial q_i/\partial\mathbf{b}\Delta\mathbf{b}$:

$$|D_1|^2 = \sum_i \left| \frac{\partial q_i}{\partial \mathbf{b}} \Delta\mathbf{b} \right|^2 = \sum_i \left| \frac{1}{\sqrt{p_i}} \frac{\partial p_i}{\partial \mathbf{b}} \Delta\mathbf{b} \right|^2 = \Delta\mathbf{b}^\mathrm{T} \sum_i \frac{1}{p_i} \frac{\partial p_i}{\partial \mathbf{b}} \frac{\partial p_i}{\partial \mathbf{b}}^\mathrm{T} \Delta\mathbf{b} = \Delta\mathbf{b}^\mathrm{T}\mathbf{F}\Delta\mathbf{b} \tag{24}$$

where the bound in *Eq (2)* is obtained from this geometric analysis.

The view of subspace projection can be extended to represent the sensitivity bound in *Eq (1)* geometrically. Intuitively, as the trace of the Fisher information is related to the surface area of a typical set given a probability density function [11], the norm of the probability gradient can be seen as a projection of this surface area as in *Figure 1* and the sensitivity bound in *Eq (1)* then follows.

### *3.2 Extended bound on the input*

Considering $\mathbf{y}$ as a deterministic function of $\mathbf{x}$, i.e. $\mathbf{y} = \mathbf{h}(\mathbf{x})$, using the law of unconscious statistician, failure probability from Eq (6) can also be written as:

$$P_\mathrm{f}(\mathbf{b}, z) = \int \mathrm{H}\left( g(\mathbf{h}(\mathbf{x})) - z \right) p(\mathbf{x} \mid \mathbf{b}) \mathrm{d}\mathbf{x} \tag{25}$$

Compared to the definition in *Eq (6)*, it can be seen that the proof given in *Section 2* still applies, except that we now have the probability sensitivity bounded by the Fisher information of the uncertain input $\mathbf{x}$ (instead of the output):

$$\left\| \frac{\partial P_\mathrm{f}(\mathbf{b}, z)}{\partial \mathbf{b}} \right\|^2 \le \mathrm{tr}\left\{ \mathbf{F}_\mathbf{x}(\mathbf{b}) \right\} \tag{26}$$

$$|\Delta P_\mathrm{f}|^2 \le \Delta\mathbf{b}^\mathrm{T}\mathbf{F}_\mathbf{x}\Delta\mathbf{b} \tag{27}$$

where $\mathbf{F}_{\mathbf{x}}$ is the Fisher Information Matrix (FIM) of the random input $\mathbf{x}$, with its $jk^{\text{th}}$ entry as:

$$F_{\mathbf{x}\,jk} = \mathbb{E}_{\mathbf{x}}\left[\frac{\partial \ln p(\mathbf{x})}{\partial b_j}\frac{\partial \ln p(\mathbf{x})}{\partial b_k}\right] \tag{28}$$

Note that in *Eqs (1)*, *(2)* and the rest of the paper, unless stated otherwise, we have implicitly assumed the FIM is with respect to the output, i.e. $\mathbf{F} = \mathbf{F}_{\mathbf{y}}$.

As the function $\mathbf{y} = \mathbf{h}(\mathbf{x})$ is often complex, a probability sensitivity bound with respect to (w.r.t) the input information, which is assumed to be known, would greatly reduce the computational cost, i.e. the sensitivity bound is obtained without evaluation of the function. In what follows, we give a brief proof that the probability sensitivity bound w.r.t the input is a weaker bound.

To link the Fisher information of the input and output, the following chain rule [16] can be used:

$$\mathbf{F}_{\mathbf{x},\mathbf{y}}(\mathbf{b}) = \mathbf{F}_{\mathbf{x}}(\mathbf{b}) + \mathbf{F}_{\mathbf{y}|\mathbf{x}}(\mathbf{b}) \tag{29}$$

and given the deterministic functional relationship $\mathbf{y} = \mathbf{h}(\mathbf{x})$

$$\mathbf{F}_{\mathbf{x},\mathbf{y}}(\mathbf{b}) = \mathbf{F}_{\mathbf{x},\mathbf{y}=\mathbf{h}(\mathbf{x})}(\mathbf{b}) = \mathbf{F}_{\mathbf{y}=\mathbf{h}(\mathbf{x})}(\mathbf{b}) \tag{30}$$

then we have the following information processing inequality for Fisher information [16]:

$$\mathbf{F}_{\mathbf{y}}(\mathbf{b}) \le \mathbf{F}_{\mathbf{x}}(\mathbf{b}) \tag{31}$$

where the inequality sign ($\le$) means the difference between the two matrices is negative definite. The inequality in *Eq (31)* follows from *Eqs (29)* and *(30)* with the fact that $\mathbf{F}_{\mathbf{y}|\mathbf{x}}(\mathbf{b})$ is a positive semidefinite (Fisher information) matrix.

Therefore, the bounds in *Eqs (1)* and *(2)* can be further extended as:

$$\left\|\frac{\partial P_{\mathrm{f}}}{\partial \mathbf{b}}\right\|^2 \le \operatorname{tr}(\mathbf{F}_{\mathbf{y}}) \le \operatorname{tr}(\mathbf{F}_{\mathbf{x}}) \tag{32}$$

$$|\Delta P_{\mathrm{f}}|^2 \le \Delta\mathbf{b}^{\mathsf{T}}\mathbf{F}_{\mathbf{y}}\Delta\mathbf{b} \le \Delta\mathbf{b}^{\mathsf{T}}\mathbf{F}_{\mathbf{x}}\Delta\mathbf{b} \tag{33}$$

The inequality relationships, with respect to both the input and output, will be examined numerically in the demonstration examples in *Section 4*.

### *3.3 Numerical implementations*

In principle, it would be possible to derive the probability sensitivities analytically if the function $\mathbf{h}(\cdot)$ is sufficiently tractable. However, most computational models are complex and can only be computed numerically. In this section, we give a brief overview of an efficient numerical method for probability sensitivities, named here as the Monte Carlo Likelihood Ratio method (MC-LR), and it will be used for some of the demonstration cases presented in Section 4. More details of the method can be found in [10,17] and a pseudo code for the algorithm used here can be found in [9].

The MC-LR method obtains estimation of the probability and its gradient w.r.t continuous parameters, e.g. $P_{\mathrm{f}}(\mathbf{b})$ and $\partial P_{\mathrm{f}}(\mathbf{b})/\partial b_j$, in a single computation run. Applying MC-LR to the failure probability defined in Eq (25):

$$\begin{aligned}\frac{\partial P_{\mathrm{f}}(\mathbf{b},z)}{\partial b_j} &= \int \mathrm{H}\left(g(\mathbf{y})-z\right)\frac{\partial \ln p(\mathbf{x}|\mathbf{b})}{\partial b_j}\,p(\mathbf{x}|\mathbf{b})\,\mathrm{d}\mathbf{x} \\ &\approx \frac{1}{N}\sum_i \mathrm{H}\left(g(\mathbf{y}_i)-z\right)\frac{\partial \ln p(\mathbf{x}_i|\mathbf{b})}{\partial b_j}\end{aligned} \tag{34}$$

where the 2nd row indicate Monte Carlo approximation, where $\mathbf{x}_i$ is a MC realisation of the random variable $\mathbf{x}$ and $N$ MC simulations are considered. Note that we estimate the sensitivity using *Eq (25)* with respect to the input $\mathbf{x}$, not the *Eq (6)*, because the input distribution is known. It is noted that although the sensitivity bounds presented in this paper apply to dependent inputs, it is assumed for simplicity in the numerical implementation below that the components of $\mathbf{x}$ are independent, that is: $p(\mathbf{x}|\mathbf{b}) = \prod_j p_j(x_j|b_{j1}, b_{j2}, \dots)$ where $p_j$ is the PDF considered for the random variable $x_j$

The MC-LR method is efficient because analytical closed-form expressions can be obtained for $\partial \ln p/\partial b_j$. For example, for Normal distribution $x \sim \mathcal{N}(\mu, \sigma^2)$:

$$\frac{\partial \ln p(x|\mu,\sigma)}{\partial \mu} = \frac{x-\mu}{\sigma^2}; \quad \frac{\partial \ln p(x|\mu,\sigma)}{\partial \sigma} = \frac{(x-\mu)^2-\sigma^2}{\sigma^3} \tag{35}$$

and for Lognormal distribution $x \sim \mathcal{LN}(\mu, \sigma^2)$:

$$\frac{\partial \ln p(x|\mu,\sigma)}{\partial \mu} = \frac{\ln x-\mu}{\sigma^2}; \quad \frac{\partial \ln p(x|\mu,\sigma)}{\partial \sigma} = \frac{(\ln x-\mu)^2-\sigma^2}{\sigma^3} \tag{36}$$

and the LR expressions for a list of commonly used distributions can be found in [18]. *Eqs (35)* and *(36)* will be used in the demonstration examples in the next section.

In some of the examples demonstrated in *Section 4* below, the Fisher Information Matrix in *Eq (5)* will also be computed using the MC-LR method. The function $\mathbf{y} = \mathbf{h}(\mathbf{x})$ is a transformation from the random variables $\mathbf{x}$ to the random variables $\mathbf{y}$, and the joint probability density function (PDF) at the output can be written as [19,20]:

$$\begin{aligned}p(\mathbf{y}|\mathbf{b}) &= \int \prod_k \delta\left[y_k - h_k(\mathbf{x})\right] p(\mathbf{x}|\mathbf{b})\,\mathrm{d}\mathbf{x} \\ &\approx \frac{1}{N}\sum_i \prod_k \delta\left[y_k - h_k(\mathbf{x}_i)\right]\end{aligned} \tag{37}$$

where $\delta(\cdot)$ is the Dirac delta function. Applying the MC-LR method to the density function above, the partial derivative w.r.t the distribution parameter $b_j$ can be computed as:

$$\begin{aligned}\frac{\partial p(\mathbf{y}\,|\,\mathbf{b})}{\partial b_j} &= \int \prod_k \delta\left[y_k - h_k(\mathbf{x})\right] \frac{\partial \ln p(\mathbf{x}\,|\,\mathbf{b})}{\partial b_j} p(\mathbf{x}\,|\,\mathbf{b})\,\mathrm{d}\mathbf{x} \\ &\approx \frac{1}{N} \sum_i \prod_k \delta\left[y_k - h_k(\mathbf{x}_i)\right] \frac{\partial \ln p(\mathbf{x}_i\,|\,\mathbf{b})}{\partial b_j}\end{aligned} \tag{38}$$

## 4 Demonstration examples

### 4.1 *Identity function $y = x$*

Consider a simple case where the random variables follow the Normal distribution, i.e. $y = x \sim \mathcal{N}(\mu, \sigma^2)$. In this case, insert the likelihood ratio expressions from *Eq (35)* into *Eq (5)* we have the FIM as (same for both $x$ and $y$):

$$\mathbf{F}(\mu,\sigma) = \begin{bmatrix} 1/\sigma^2 & 0 \\ 0 & 2/\sigma^2 \end{bmatrix} \tag{39}$$

We consider in this case the probability measure to be the same as the cumulative distribution function (CDF), i.e. $P_\mathrm{f}(y) = P(Y \le y)$. For Normal distribution, we have:

$$P_\mathrm{f}(y) = \frac{1}{2}\left[1 + \mathrm{erf}\left(\frac{y-\mu}{\sqrt{2}\sigma}\right)\right] \tag{40}$$

and its partial derivative with respect to the distribution parameters are:

$$\begin{aligned}\frac{\partial P_\mathrm{f}(y)}{\partial \mu} &= -\frac{1}{\sqrt{2\pi}\sigma}\exp\left(-\left(\frac{y-\mu}{\sqrt{2}\sigma}\right)^2\right) = -p(y) \\ \frac{\partial P_\mathrm{f}(y)}{\partial \sigma} &= -\frac{y-\mu}{\sigma}\frac{1}{\sqrt{2\pi}\sigma}\exp\left(-\left(\frac{y-\mu}{\sqrt{2}\sigma}\right)^2\right) = -\frac{y-\mu}{\sigma}p(y)\end{aligned} \tag{41}$$

where $p(y)$ is the PDF for the Normal distributed random variable. The norm of the probability sensitivity is then:

$$\left\|\frac{\partial P_\mathrm{f}}{\partial \mathbf{b}}\right\|^2 = \left(\frac{\partial P_\mathrm{f}}{\partial \mu}\right)^2 + \left(\frac{\partial P_\mathrm{f}}{\partial \sigma}\right)^2 = p^2(y)\left[1 + \left(\frac{y-\mu}{\sigma}\right)^2\right] \tag{42}$$

Taking the 1st derivative of *Eq (42)* w.r.t $y$ and looking for the stationary points, the maximum value of the sensitivity norm can be found. As shown in *Appendix B*, it turns out that in this case, both the 1st and 2nd derivatives of the function are zero at $y = \mu$ and there is a flat region around that point. As the exponential function has a bigger growth rate than the quadratic function, the asymptotes of the function tends to zero away from $\mu$ (see also the numerical results in Figure 2 with $\mu = 1$ and $\sigma = 0.2$). The probability sensitivity at $y = \mu$ is

$$\left\| \frac{\partial P_{\mathrm{f}}(y=\mu)}{\partial \mathbf{b}} \right\|^2 = \frac{1}{2\pi\sigma^2} < \frac{3}{\sigma^2} = \mathrm{tr}(\mathbf{F}) \tag{43}$$

and the bound in *Eq (1)* is clearly satisfied as compared to the trace of the FIM in *Eq (39)*

In addition to the analytical results, *Figure 2* also shows the Fisher information and the probability sensitivity computed using the MC-LR method described in *Section 3.3*, and good agreement is obtained.

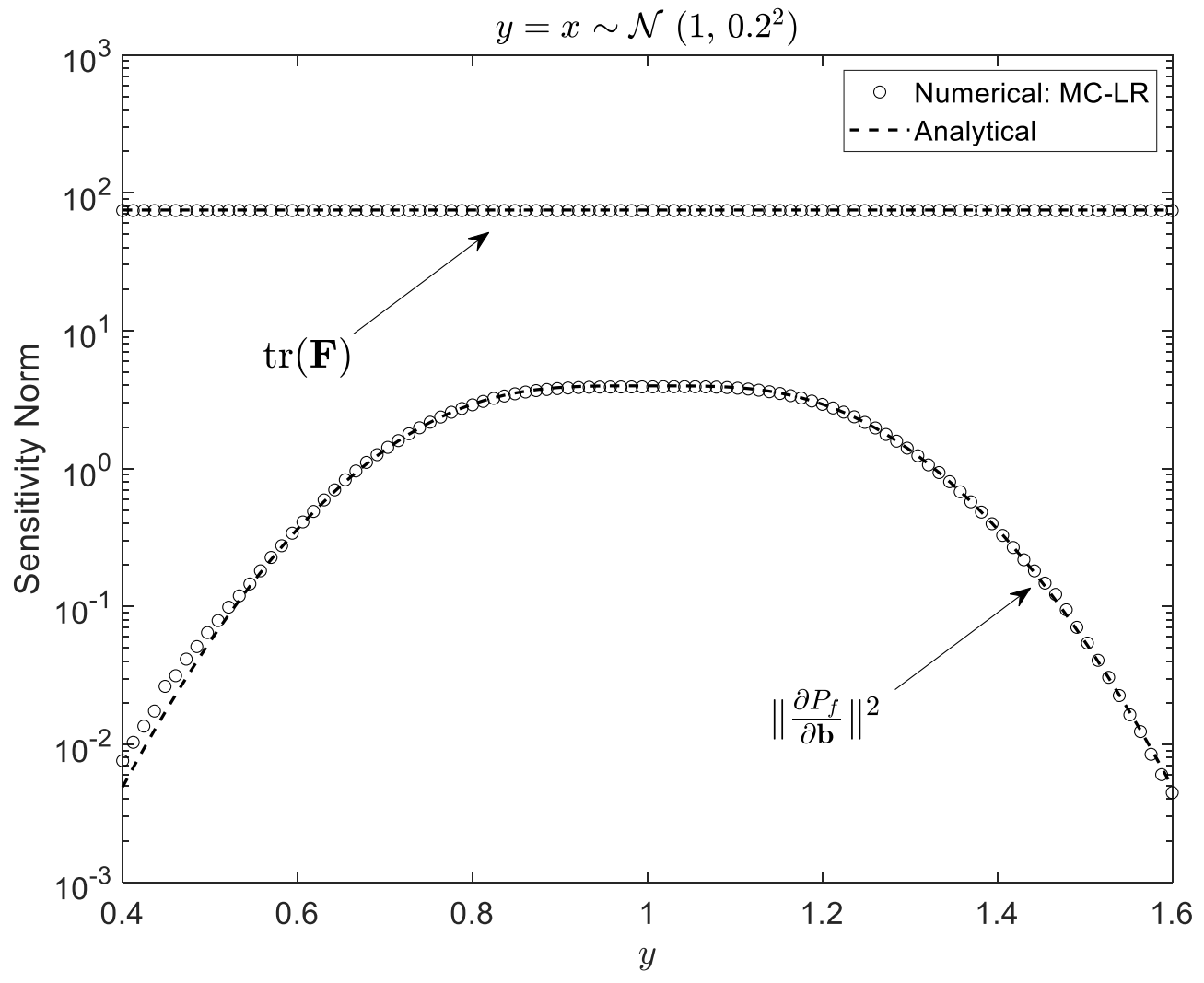

*Figure 2 benchmark case with a trivial identity function $y = x$*

## 4.2 A simple harmonic oscillator

Consider small amplitude vibration of a simple harmonic oscillator (SHO) subject to a harmonic driving force, its frequency response function can be described non-dimensionally as [21]:

$$H = \left[1 - \beta^2 + i2\zeta\beta\right]^{-1} \tag{44}$$

where $\beta = \omega/\omega_n$ is the ratio between the forcing frequency $\omega$ and the natural frequency of the SHO $\omega_n$, and $\zeta$ is the non-dimensional viscous damping factor. In this example, both $\beta$ and $\zeta$ are considered to be random and follow the Normal distribution: $\beta \sim \mathcal{N}(1, 0.1^2)$ and $\zeta \sim \mathcal{N}(0.1, 0.01^2)$. The FIM for the input can be computed using *Eq (39)* analytically, however, the probability sensitivity and the Fisher information of the output need to be estimated numerically using the MC-LR method described in *Section 3.3*. Like the previous case, the probability measure in this case is simply the cumulative distribution function of the response. The results are shown in *Figure 3*, where the threshold for the probability sensitivity is defined as the $z^{\text{th}}$ percentile of the random output. It can be seen that the bounds are clearly satisfied, with the probability sensitivity bound w.r.t the input, $\mathrm{tr}(\mathbf{F}_\mathbf{x})$, as the weaker bound.

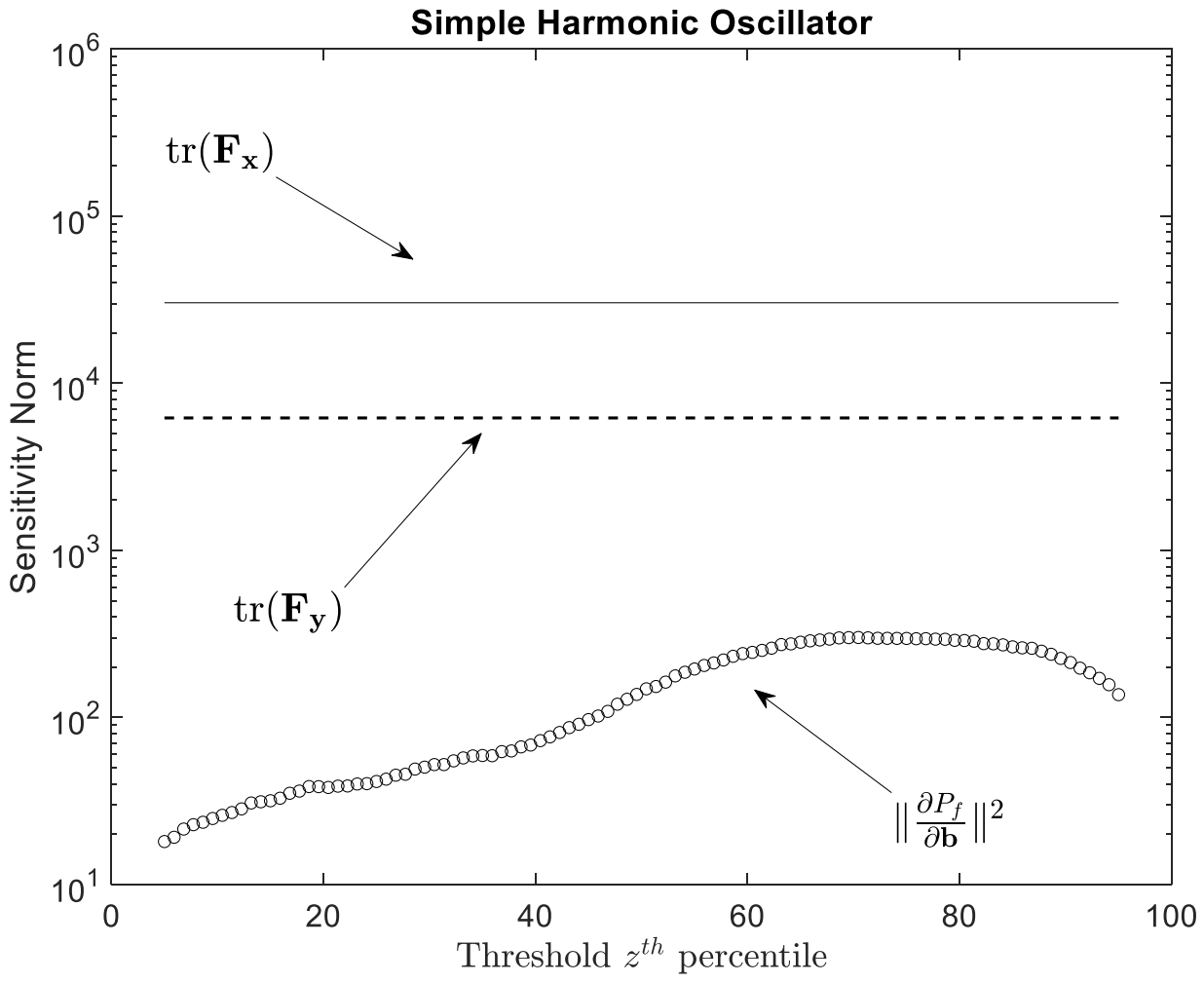

*Figure 3 A simple case with a simple harmonic oscillator. Results obtained numerically using MC-LR*

### 4.3 *A thin cantilever beam*

While previous examples have only scalar outputs, in this example, we consider a vector output using a thin cantilever beam, as shown in *Figure 4 (a)*. The cantilever beam is subject to a bandlimited white noise excitation, where only the first three modes are excited, at the middle span position. In this case, the outputs of interest are the peak r.m.s acceleration and strain responses along the beam (output $\mathbf{y}$ is 2-dimensional). The frequency response functions for both acceleration and strain responses, at different positions along the beam, are obtained via modal summation and the modal damping is assumed to be 0.1 for all modes, see *Appendix C* for details.

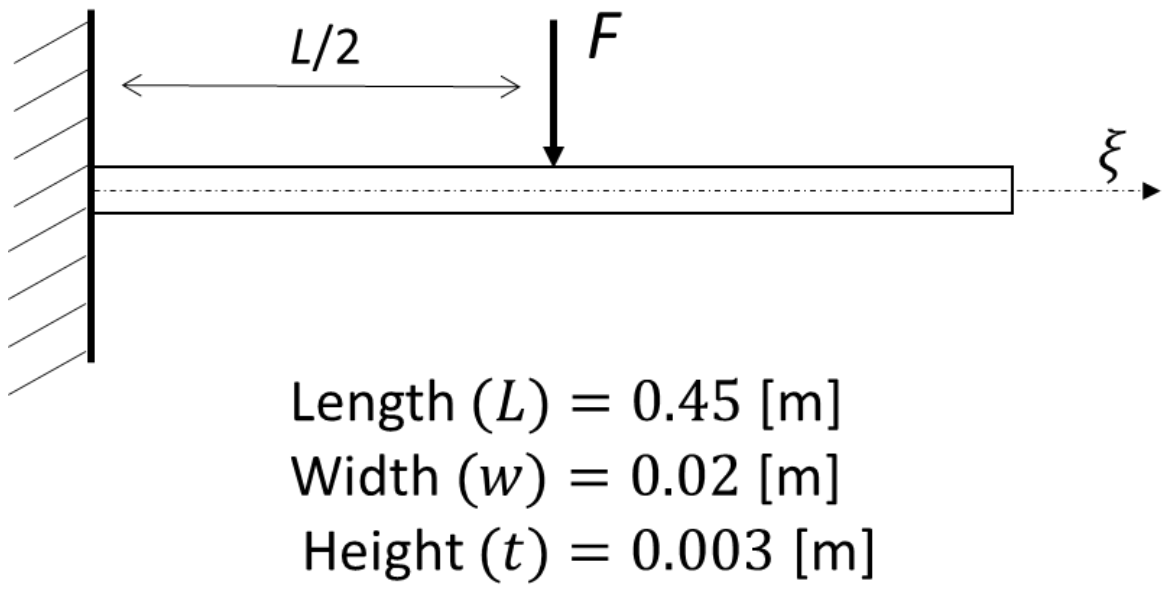

*(a) A cantilever beam subject to white noise excitation of unit amplitude*

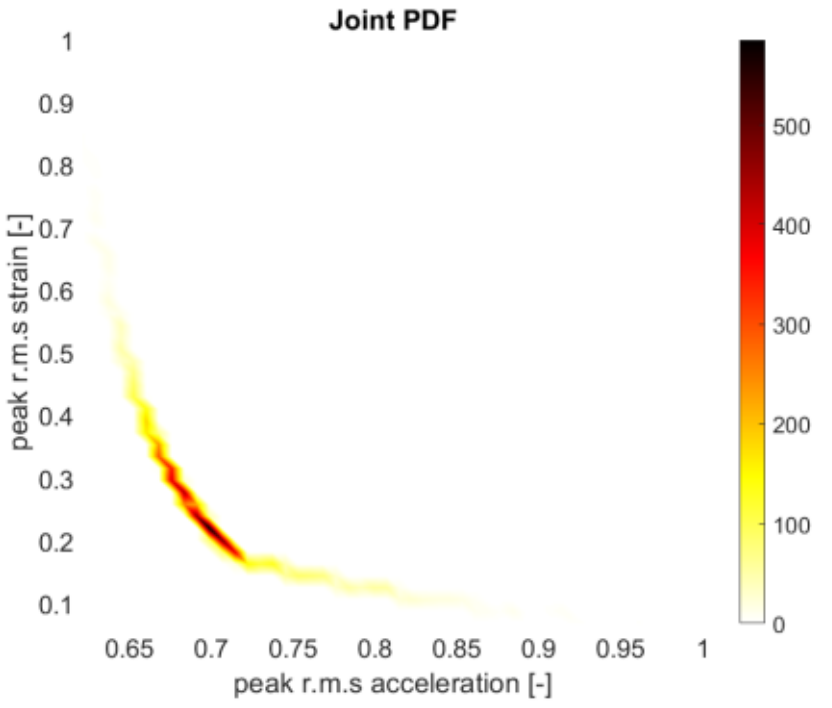

*(b) joint PDF of the peak r.m.s response along the beam*

*Figure 4 A thin cantilever beam case. (a) parameters for the beam; (b) the joint PDF of the 2-dimensional responses consisting of peak r.m.s acceleration and strain response. 'peak' indicates the maximum response along the beam for each sample of the random input. The two type of responses are normalised by the maximum values across the ensemble of the random samples.*

For the input, both Young's modulus $E$ and material density $\rho$ are considered to be random and follow a Lognormal distribution: $E \sim \mathcal{LN}(24.85, 0.47^2)$ and $\rho \sim \mathcal{LN}(7.88, 0.2^2)$. This gives the mean value of 69

GPa and $2700\ \mathrm{kg/m^3}$ for the Young's modulus and material density, i.e. an aluminium beam (nominally). In this case, the output $\mathbf{y}$ is two dimensional and the joint PDF of the (normalised) acceleration and strain responses are shown in *Figure 4 (b)*. Different from previous cases where the probability measure is simply the cumulative distribution function, it is assumed here that the performance function $g(\cdot)$ in *Eq (6)* is a square function of the peak responses along the beam as:

$$g(y) = \max^2\left\{y_{\mathrm{acc}}(\xi)\right\} + \max^2\left\{y_{\mathrm{str}}(\xi)\right\} \tag{45}$$

where $y_{\mathrm{acc}}$ and $y_{\mathrm{str}}$ are the acceleration and strain responses respectively. The probability sensitivity and the Fisher information of the output, using the joint PDF from *Figure 4 (b)*, are then evaluated using MC-LR method described in *Section 3.3*. The results are shown in *Figure 5* with the corresponding lines clearly labelled as previous cases. The Fisher information of the inputs, which follow Lognormal distribution, can be calculated analytically using *Eq (39)* similar to the Normal distribution case. It can be seen from Figure 5 that the bounds are clearly satisfied. As the probability sensitivity depends on the specific threshold $z$ and the performance function or failure mode $g(\cdot)$ given in *Eq (45)*, which can vary during a decision process, the Fisher information potentially provides a more robust guidance for decision making under uncertainties.

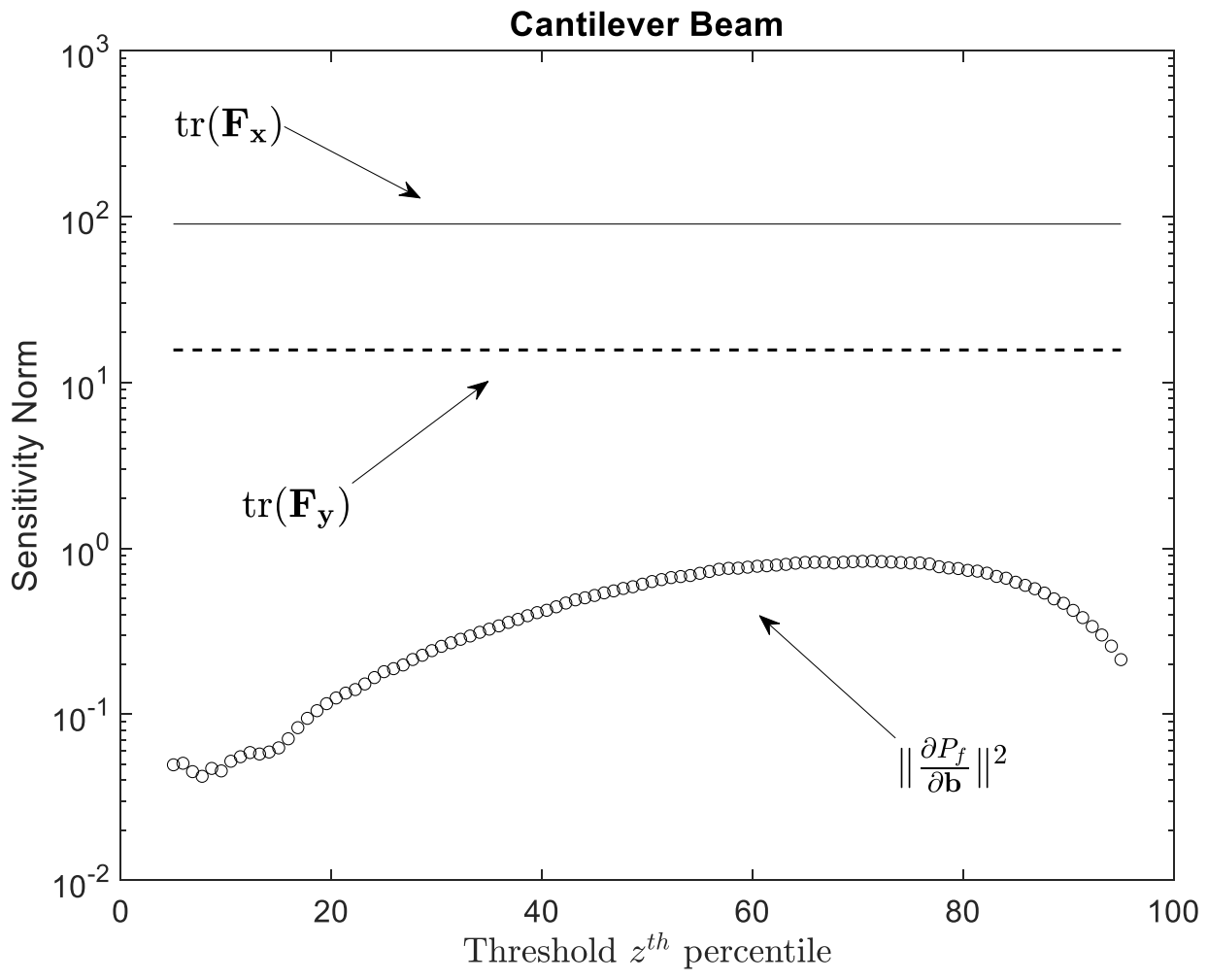

*Figure 5 Sensitivity results for a cantilever beam. Results obtained numerically using MC-LR*

## 5 Conclusions

A new upper bound for probabilistic sensitivities, using the Fisher information and the Kullback-Leibler divergence, has been presented with mathematical proofs. Contrasting to the probability sensitivity metric, the Fisher information is independent of the thresholds and the failure modes on which the probability measure is based and thus allows inclusion of wider uncertainties in decision making. Recent studies have shown that these two metrics are different but closely related. The inequality derived in this paper thus provide a solid mathematical basis to decisions based on these sensitivity metrics.

Using the information processing inequality, the sensitivity bound for the probability has been extended to the Fisher information of both the uncertain input and output. The Fisher information of the input is shown as the weaker bound. The extended inequality has been demonstrated with a few numerical examples, where an efficient Monte Carlo based Likelihood Ratio method has been applied for the sensitivity computation.

The proof presented in Section 2 makes no assumption of the functional form of the probability measure. Nevertheless, in the simple but common case where the probability of interest is simply the (complementary) cumulative distribution function, the bound for the perturbation of the probability by the K-L divergence is found to be a consequence of the well-known Pinsker's inequality.

In addition to decision-making in the presence of uncertainties, future research will consider the potential application of the bound on the gradient of the probability function to stochastic optimizations where the objective function are defined probabilistically. Moreover, the applications of which the function $h(\cdot)$ is stochastic will be explored.

## Acknowledgement

This work has been funded by the Engineering and Physical Sciences Research Council through the award of a Programme Grant "Digital Twins for Improved Dynamic Design", Grant No. EP/R006768. For the purpose of open access, the author has applied a Creative Commons Attribution (CC BY) licence to any Author Accepted Manuscript version arising. The author is grateful to Professor Robin Langley, University of Cambridge, for advices of using information-theoretic metrics and comments on an early draft of the work.

## Data availability statement

The datasets generated during and/or analysed during the current study are available in the GitHub repository: http://doi.org/10.5281/zenodo.6615192.

## Appendix A Relative entropy approximation

The perturbation of entropy, quantified using K-L divergence, can be represented as a quadratic form using the positive semidefinite symmetric Fisher Information Matrix (FIM)

$$\begin{aligned}
\Delta H &\equiv KL\left[ p(\mathbf{y}\,|\,\mathbf{b}) \,\|\, p(\mathbf{y}\,|\,\mathbf{b}+\Delta\mathbf{b}) \right] \\
&= \int p(\mathbf{y}\,|\,\mathbf{b}) \ln\left[ \frac{p(\mathbf{y}\,|\,\mathbf{b})}{p(\mathbf{y}\,|\,\mathbf{b}+\Delta\mathbf{b})} \right] \mathrm{d}\mathbf{y} \\
&\approx \int p \ln p \mathrm{d}\mathbf{y} - \int p \ln p \mathrm{d}\mathbf{y} - \Delta\mathbf{b}^{\mathsf{T}} \int p \nabla \ln p \mathrm{d}\mathbf{y} - \frac{1}{2} \Delta\mathbf{b}^{\mathsf{T}} \int p \nabla^2 \ln p \mathrm{d}\mathbf{y} \Delta\mathbf{b} \\
&= \Delta\mathbf{b}^{\mathsf{T}} \int p \frac{1}{p} \nabla \ln p \mathrm{d}\mathbf{y} - \frac{1}{2} \Delta\mathbf{b}^{\mathsf{T}} \int p \left[ \frac{1}{p} \nabla^2 p - \frac{1}{p^2} \nabla p^{\mathsf{T}} \nabla p \right] \mathrm{d}\mathbf{y} \Delta\mathbf{b} \\
&= \frac{1}{2} \Delta\mathbf{b}^{\mathsf{T}} \int \frac{1}{p} \nabla p^{\mathsf{T}} \nabla p \mathrm{d}\mathbf{y} \Delta\mathbf{b} \\
&= \frac{1}{2} \Delta\mathbf{b}^{\mathsf{T}} \mathbf{F} \Delta\mathbf{b}
\end{aligned} \tag{A1}$$

where $\nabla, \nabla^2$ represent the gradient vector and Hessian matrix respectively. Note that the gradient vector is denoted as a column vector. In the 3rd line of the equation above, the Taylor expansion of the perturbed probability at the point **b** is taken:

$$\ln\left[ p\left(\mathbf{b}+\Delta\mathbf{b}\right) \right] = \ln p\left(\mathbf{b}\right) + \Delta\mathbf{b}^{\mathsf{T}} \left\{ \nabla \ln p\left(\mathbf{b}\right) \right\} + \frac{1}{2} \Delta\mathbf{b}^{\mathsf{T}} \left\{ \nabla^2 \ln p\left(\mathbf{b}\right) \right\} \Delta\mathbf{b} + O\left(\Delta\mathbf{b}^3\right) \tag{A2}$$

where $p(\mathbf{b} + \Delta\mathbf{b})$ is the simplified notation for $p(\mathbf{y}|\mathbf{b} + \Delta\mathbf{b})$ and third and higher order terms are ignored.

Although the K-L divergence is asymmetric in general, it can be shown the expression in *Eq (A1)* is still valid even if the perturbation order is reversed.

$$
\begin{aligned}
\Delta H &\equiv KL\left[p(\mathbf{y}\,|\,\mathbf{b}+\Delta\mathbf{b})\,\|\,p(\mathbf{y}\,|\,\mathbf{b})\right] \\
&= \int p(\mathbf{y}\,|\,\mathbf{b}+\Delta\mathbf{b})\ln\left[\frac{p(\mathbf{y}\,|\,\mathbf{b}+\Delta\mathbf{b})}{p(\mathbf{y}\,|\,\mathbf{b})}\right]\mathrm{d}\mathbf{y} \\
&\approx \int\left[p+\Delta\mathbf{b}^{\mathsf{T}}\nabla p+\frac{1}{2}\Delta\mathbf{b}^{\mathsf{T}}\nabla^2 p\Delta\mathbf{b}\right]\left[\ln p+\Delta\mathbf{b}^{\mathsf{T}}\nabla\ln p+\frac{1}{2}\Delta\mathbf{b}^{\mathsf{T}}\nabla^2\ln p\Delta\mathbf{b}-\ln p\right]\mathrm{d}\mathbf{y} \\
&\approx \int\left[p\Delta\mathbf{b}^{\mathsf{T}}\nabla\ln p+\frac{1}{2}p\Delta\mathbf{b}^{\mathsf{T}}\nabla^2\ln p\Delta\mathbf{b}+\frac{1}{p}\Delta\mathbf{b}^{\mathsf{T}}\nabla\ln p\left(\nabla\ln p\right)^{\mathsf{T}}\Delta\mathbf{b}\right]\mathrm{d}\mathbf{y} \\
&= \Delta\mathbf{b}^{\mathsf{T}}\int\nabla p\,\mathrm{d}y+\frac{1}{2}\Delta\mathbf{b}^{\mathsf{T}}\mathbb{E}\left[\nabla^2\ln p\right]\Delta\mathbf{b}+\Delta b^2\mathbb{E}\left[\nabla\ln p\left(\nabla\ln p\right)^{\mathsf{T}}\right]\Delta\mathbf{b} \\
&= \frac{1}{2}\Delta\mathbf{b}^{\mathsf{T}}\mathbf{F}\Delta\mathbf{b}
\end{aligned}
\tag{A3}
$$

where the 2nd term at the second to last steps is just $-1/2\mathbf{F}$:

$$
\mathbb{E}\left[\nabla^2\ln p\right]=\mathbb{E}\left[\frac{1}{p}\nabla^2 p-\frac{1}{p^2}\nabla p\left(\nabla p\right)^{\mathsf{T}}\right]=\mathbb{E}\left[\frac{1}{p}\nabla^2 p\right]-\mathbb{E}\left[\nabla\ln p\left(\nabla\ln p\right)^{\mathsf{T}}\right]=-\mathbf{F}
\tag{A4}
$$

## Appendix B Derivatives for case 1

$$
\left\|\frac{\partial P_{\mathrm{f}}}{\partial\mathbf{b}}\right\|^2=\left(\frac{\partial P_{\mathrm{f}}}{\partial\mu}\right)^2+\left(\frac{\partial P_{\mathrm{f}}}{\partial\sigma}\right)^2=p^2\left(y\right)\left[1+\left(\frac{y-\mu}{\sigma}\right)^2\right]
\tag{B1}
$$

Take the first derivative:

$$
\begin{aligned}
\frac{\partial}{\partial y}\left\|\partial P_{\mathrm{f}}/\partial\mathbf{b}\right\|^2 &= 2p\left(y\right)\frac{\partial p\left(y\right)}{\partial z}\left[1+\left(\frac{y-\mu}{\sigma}\right)^2\right]+2p^2\left(y\right)\frac{y-\mu}{\sigma^2} \\
&= -2p^2\left(y\right)\frac{\left(y-\mu\right)^3}{\sigma^4}
\end{aligned}
\tag{B2}
$$

where $\frac{\partial p}{\partial y}=-\frac{y-\mu}{\sigma^2}p$ is used.

$$
\begin{aligned}
\frac{\partial^2}{\partial y^2}\left\|\partial P_{\mathrm{f}}/\partial\mathbf{b}\right\|^2 &= -\frac{1}{\sigma^4}\left[2p\frac{\partial p}{\partial z}\left(y-\mu\right)^3+3p^2\left(y-\mu\right)^2\right] \\
&= -\frac{1}{\sigma^4}\left[-\frac{1}{\sigma^2}2p^2\left(y-\mu\right)^4+3p^2\left(y-\mu\right)^2\right]
\end{aligned}
\tag{B3}
$$

It can be seen that both the 1st and 2nd derivatives are zero when $y=\mu$.

## Appendix C Vibration of a thin cantilever beam

The mode shape for a thin cantilever beam is [22]:

$$\phi_r(\xi) = \frac{C}{\sin\beta_r L - \sinh\beta_r L}\Big[\big(\sin\beta_r L - \sinh\beta_r L\big)\big(\sin\beta_r\xi - \sinh\beta_r\xi\big) + \big(\cos\beta_r L + \cosh\beta_r L\big)\big(\cos\beta_r\xi - \cosh\beta_r\xi\big)\Big] \tag{C1}$$

where $\beta_r, r = 1, 2, \ldots$ is the solution to the characteristic equation below:

$$\cos\beta L \cosh\beta L = -1 \tag{C2}$$

Solution of *Eq* can be obtained numerically, e.g. Newton Raphson method, yielding an infinite set of $\beta_r$. It is assumed in the case study considered, only the first three modes are excited by the bandlimited white noise and the first three solutions of Eq are $\beta_1 L = 1.875$, $\beta_2 L = 4.694$ and $\beta_3 L = 7.855$.

The displacement frequency response function using modal summation is:

$$H_{\text{dis}}(\xi_{\text{re}}, \xi_{\text{ex}}, \omega) = \sum_{r=1} \frac{\phi_r(\xi_{\text{re}})\phi_r(\xi_{\text{ex}})}{\omega_r^2 - \omega^2 + 2i\zeta_r\omega_r\omega} \tag{C3}$$

where $\xi_{\text{re}}$ is the coordinate for the response measurement and $\xi_{\text{ex}}$ indicates the forcing position. $\zeta_r$ is the modal damping and it is assumed to be 0.1 for all modes and $\omega_r$ is the $r^{th}$ natural frequency given by:

$$\omega_r = (\beta_r L)^2 \sqrt{\frac{EI}{mL^4}} \tag{C4}$$

where $E$ is the Young's modulus, $I$ is the area moment of inertia and $m$ is the mass per unit length.

Similarly, the frequency response function for strain is:

$$H_{\text{str}}(\xi_{\text{re}}, \xi_{\text{ex}}, \omega) = \sum_{r=1} \frac{\phi_r''(\xi_{\text{re}})\phi_r(\xi_{\text{ex}})}{\omega_r^2 - \omega^2 + 2i\zeta_r\omega_r\omega} \tag{C5}$$

Considering a white noise excitation $S_0$, the acceleration and strain response spectrum is given

$$S_{\text{acc}}(\xi_{\text{re}}, \xi_{\text{ex}}, \omega) = \left|\omega^2 H_{\text{dis}}\right|^2 S_0(\xi_{\text{ex}}) \tag{C6}$$

$$S_{\text{str}}(\xi_{\text{re}}, \xi_{\text{ex}}, \omega) = \left|H_{\text{str}}\right|^2 S_0(\xi_{\text{ex}}) \tag{C7}$$

The r.m.s response, for both acceleration and strain, can then be obtained as [21]:

$$y(\xi) = \sqrt{\sum 2S(\xi, \omega)\,\mathrm{d}\omega} \tag{C8}$$